\documentclass[fleqn,12pt]{article}
\usepackage{mathptmx}
\usepackage{esp}
\usepackage{epsfig,epsf}
\usepackage{rotate}
\usepackage{graphics}
\usepackage{graphicx}
\usepackage[figuresright]{rotating}

\newcommand{\AmS}{{\protect\the\textfont2
  A\kern-.1667em\lower.5ex\hbox{M}\kern-.125emS}}

\title{Ultracold Scattering Processes in Three-Atomic Helium Systems\thanks{Published
in \htmladdnormallink{Nucl. Phys. A \textbf{790} (2007), 752c--756c.}
{http://dx.doi.org/10.1016/j.nuclphysa.2007.03.120}
}
\thanks{This
work was supported by the Deu\-t\-s\-che
For\-s\-ch\-ungs\-ge\-mein\-schaft (DFG), the Heisenberg-Landau
Program, Alexander von Humboldt Foundation, and the Russian
Foundation for Basic Research.}}

\author{E. A. Kolganova\address[MCSD]{BLTP, JINR, Joliot-Curie 6,
141980 Dubna, Moscow Region, Russia}\address[PI]{Physikalisches Institut, Universit\"at Bonn,
Endenicher Allee 11-13, D-53115 Bonn, Germany},
        A. K. Motovilov\addressmark[MCSD]
        and
        W. Sandhas\addressmark[PI]
        }

\begin{document}

% typeset front matter
\maketitle

\begin{abstract}
We review results on scattering observables for $^4$He--$^4$He$_2$
and $^3$He--$^4$He$_2$ collisions. We also study the effect of
varying the coupling constant of the atom-atom interaction on the
scattering length.
\end{abstract}

\section{INTRODUCTION}

Experimentally, $^4$He dimers have been observed in 1993 by Luo {\em
at al.} \cite{DimerExp}, and in 1994 by Sch\"ollkopf and Toennies
\cite{Science}. In the latter investigation the existence of $^4$He
trimers has also been demonstrated. Later on, Grisenti {\em et al.}
\cite{exp} measured a bond length of $52 \pm 4$ {\AA} for
$^4$He$_2$, which indicates that this dimer is the largest known
diatomic molecular ground state. Based on this measurement they
estimated a scattering length of $104^{+8}_{-18}$ {\AA} and a dimer
energy of $1.1^{+0.3}_{-0.2}$ mK \cite{exp}. Further investigations
concerning $^4$He trimers and tetramers have been reported in
\cite{Toennies2002}, but with no results on size and binding
energies. To the best of our knowledge, we are not aware of any
experimental results on binding energy and size of the
asymmetric helium trimer consisting of two $^4$He and one $^3$He
atoms.

Many theoretical calculations of the helium three-atomic systems
were performed in the past for various interatomic potentials.
Variational, hyperspherical, and Faddeev-type techniques have been
employed in this context \cite{Cornelius}--\cite{Roudnev}. It was
found that the $^4$He trimer has two bound states of total angular
momentum zero: a ground state of about $126$ mK and an excited
state of about $2.28$ mK. Experimentally this Efimov-type excited
state has not yet been observed. It should be mentioned, however,
that the year 2006 is noticeable due the first convincing
experimental evidence for the Efimov effect in an ultracold gas of
caesium atoms~\cite{Kraemer}.

Due to the smaller mass of the $^3$He atom, the $^3$He--$^4$He
system is unbound. Nevertheless, the $^3$He$^4$He$_2$ trimer exists,
though with a binding energy of about 14 mK. In contrast to the
symmetric case, there is no excited state in the asymmetric
$^3$He$^4$He$_2$ system.

Phase shifts of $^4$He--$^4$He$_2$ elastic scattering at ultra-low
energies have been calculated for the first time in
\cite{PRA-CPL,MSSK} both below and above the three-body threshold.
An alternative \textit{ab initio} calculation was performed in
\cite{Roudnev1}, but only below the threshold. The only available
results on the $^3$He--$^4$He$_2$ phase shifts were obtained in
\cite{FBS2004}. For completeness we notice that zero-range models
are able to reproduce the $^4$He--$^4$He$_2$ scattering situation
\cite{BraatenHammer}.

In what follows we present our results on binding energies and
scattering observables in the helium three-atomic systems obtained
mainly with the LM2M2 potential by Aziz and Slaman \cite{Aziz91}.

\section{RESULTS}

In our calculations we employed the hard-core version of the Faddeev
differential equations developed in \cite{Vestnik}. Table
\ref{tableTrimers} summarizes trimer binding energies and
He--He$_2$ scattering lengths, calculated with the LM2M2 potential.
The binding energies of the $^4$He trimer ground state ($E_{^4{\rm
He}_3}$) and exited state ($E_{^4{\rm He}_3}^{*}$) are presented in
the first two rows. These results demonstrate the good agreement between
the different methods.

\begin{table}[h]
\caption {Results for binding energies of the $^4\mathrm{He}_3$ and
$^3{\rm He}^4{\rm He}_2$ trimers and the He--He$_2$ scattering lengths.}
\label{tableTrimers}
\begin{tabular}{cccccccc}
\hline \\[-2ex]
            & \cite{MSSK}$^\mathrm{a}$ & \cite{Nielsen} & \cite{BlumeGreene} & \cite{RoudnevYak} &  \cite{Bressanini}
&\  \cite{Barletta}  &\ \cite{Lazaus}  \\[0.1ex]
\hline \\[-2.5ex]
$-E_{^4{\rm He}_3}$ (mK) &\  $125.9$ &\ $125.2$  &\ $125.5$ &\ 126.41  &\ 126.39  &\   126.4  & $126.39$\\[1ex]
$-E_{^4{\rm He}_3}^{*}$ (mK) &\  $2.276^\mathrm{b}$ &\ $2.269$  &\ $2.19$ &\  2.271  &\    &\  2.265  & $2.268$\\[1ex]
$\ell^{(^4{\rm He}-^4{\rm He}_2)}_{\rm sc}$ (\AA)
&\  $115.5^\mathrm{c}$ &\   &\ $126.$ &\  115.4$^\mathrm{d}$   &\  &\  & $115.2$\\[0.5ex]
\hline
\\[-2.5ex]
$-E_{^3{\rm He}^4{\rm He}_2}$ (mK) &\  $13.84^\mathrm{e}$ &\ $13.66$  &\  &\ 14.4$^\mathrm{f}$    &\ 14.165  &\ & \\[1ex]
$\ell^{(^3{\rm He}-^4{\rm He}_2)}_\mathrm{sc}$ (\AA) &\  $21.0^\mathrm{e}$ &\   &\ &\  19.3$^\mathrm{f}$   &\   &\  & \\[0.5ex]
\hline
\end{tabular}
\\[2pt]
{\footnotesize
$^\mathrm{a}$Calculations with maximal value $l_\mathrm{max}=4$ of the subsystem
angular momentum. %\\
$^\mathrm{b}$This value was rounded in \cite{MSSK}. \\
$^\mathrm{c}$Result of extrapolation, see \cite{PRA04}. %\\
$^\mathrm{d}$Result from Ref. \cite{Roudnev1}. %\\
$^\mathrm{e}$Result from Ref. \cite{FBS2004}. %\\
$^\mathrm{f}$Result from Ref. \cite{Roudnev}.}
\vspace*{-0.5cm}
\end{table}

\begin{table}[htb]
\caption{ Dependence of the $^4$He dimer and trimer energies (mK)
and the $^4$He--$^4$He and $^4$He--$^4$He$_2$ scattering lengths
(\AA) on the potential strength $\lambda$. The three-body results
were obtained with a ma\-xi\-mal value $l_\mathrm{max}=0$ of the subsystem
angular momentum. The HFD-B potential from Ref. \cite{Aziz87} was used.}
\label{tableScLength1}
%\beforetab
\centering{
\begin{tabular}{lcccccc}
\hline $\lambda$ & $\epsilon_d$ & $\epsilon_d - E^*$ &
$\epsilon_d - E_{\rm virt}$ & $\epsilon_d - E^{**}$ &
$\ell^{(1+2)}_{\rm sc}$ & $\ell^{(1+1)}_{\rm sc}$   \\
\hline
1.30 & $-199.45$ & - & 1.831 & - & $-61$ & 11.4  \\
1.20 & $-99.068$ & - & 0.01552 & - & $-340$ & 14.7  \\
1.18 & $-82.927$ & - & 0.00058 & - & $-1783$ & 15.8  \\
1.17 & $-75.367$ & 0.0063 & - & - & 8502 & 16.3  \\
1.15 & $-61.280$ & 0.0737 & - & - & 256 & 17.7  \\
1.10 & $-32.222$ & 0.4499 & - & - & 152 & 23.1  \\
1.0 & $-1.685$ & 0.773 & - & - & 160 & 88.6  \\
0.995 & $-1.160$ & 0.710 & - & - & 151 & 106  \\
0.990 & $-0.732$ & 0.622 & - & - & 143 & 132 \\
0.9875$\quad$ & $-0.555$ & 0.222 & - & - & 125 & 151  \\
0.985 & $-0.402$ & 0.518 & 0.097 & - & 69 & 177  \\
0.982 & $-0.251$ & 0.447 & 0.022 & - & $-75$ & 223  \\
0.980 & $-0.170$ & 0.396 & 0.009 & - & $-337$ & 271 \\
0.9775 & $-0.091$ & 0.328 & 0.003& - & $-6972$ & 370 \\
0.975 & $-0.036$ & 0.259 & - & 0.002 & 7120 & 583 \\
0.973 & $-0.010$ & 0.204 & - & 0.006 & 4260 & 1092 \\
\hline
\end{tabular}}
\vspace{-0.5cm}
\end{table}

%%%%%%%%%%%%%%%%%%%%

\begin{figure}[h]
%[htb]
\begin{minipage}[t]{75mm}
\vspace*{-0.2truecm}

\hspace*{-0.4truecm}{\includegraphics[angle=-90,width=8.cm]{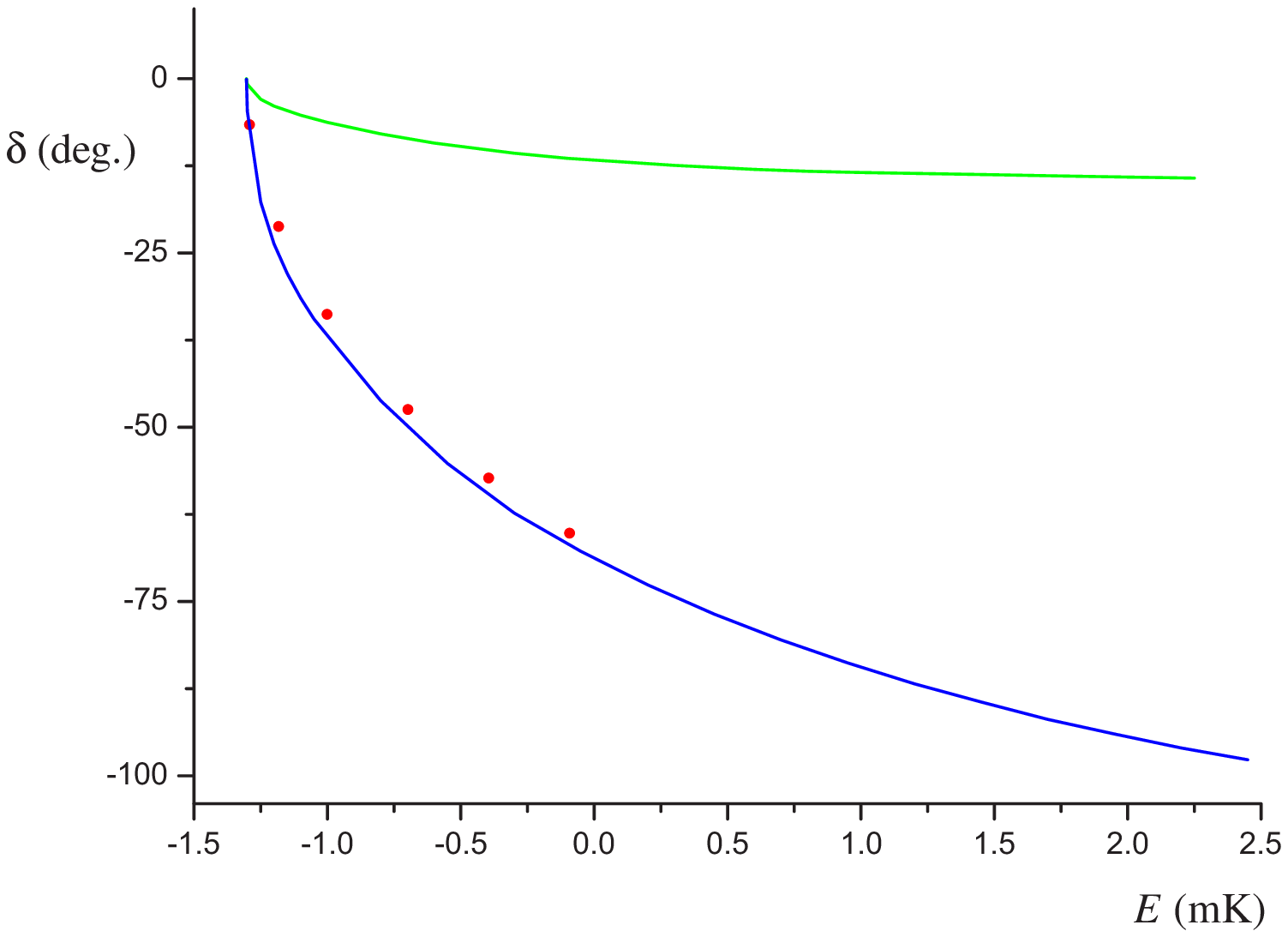}}
\vspace*{-1.05truecm}

\caption{S-wave He--He$_2$ phase shifts $\delta(E)$ obtained with
the LM2M2 potential. The lower curve depicts  the $^4$He--$^4$He$_2$
results of \cite{MSSK}, the dots are the ones found in
\cite{Roudnev1}. The upper curve represents the $^3$He--$^4$He$_2$
phase shifts of Ref. \cite{FBS2004}.} \label{Fig-phases}
\end{minipage}
\hspace{\fill}
\begin{minipage}[t]{80mm}
{\includegraphics[angle=-90,width=7.8cm]{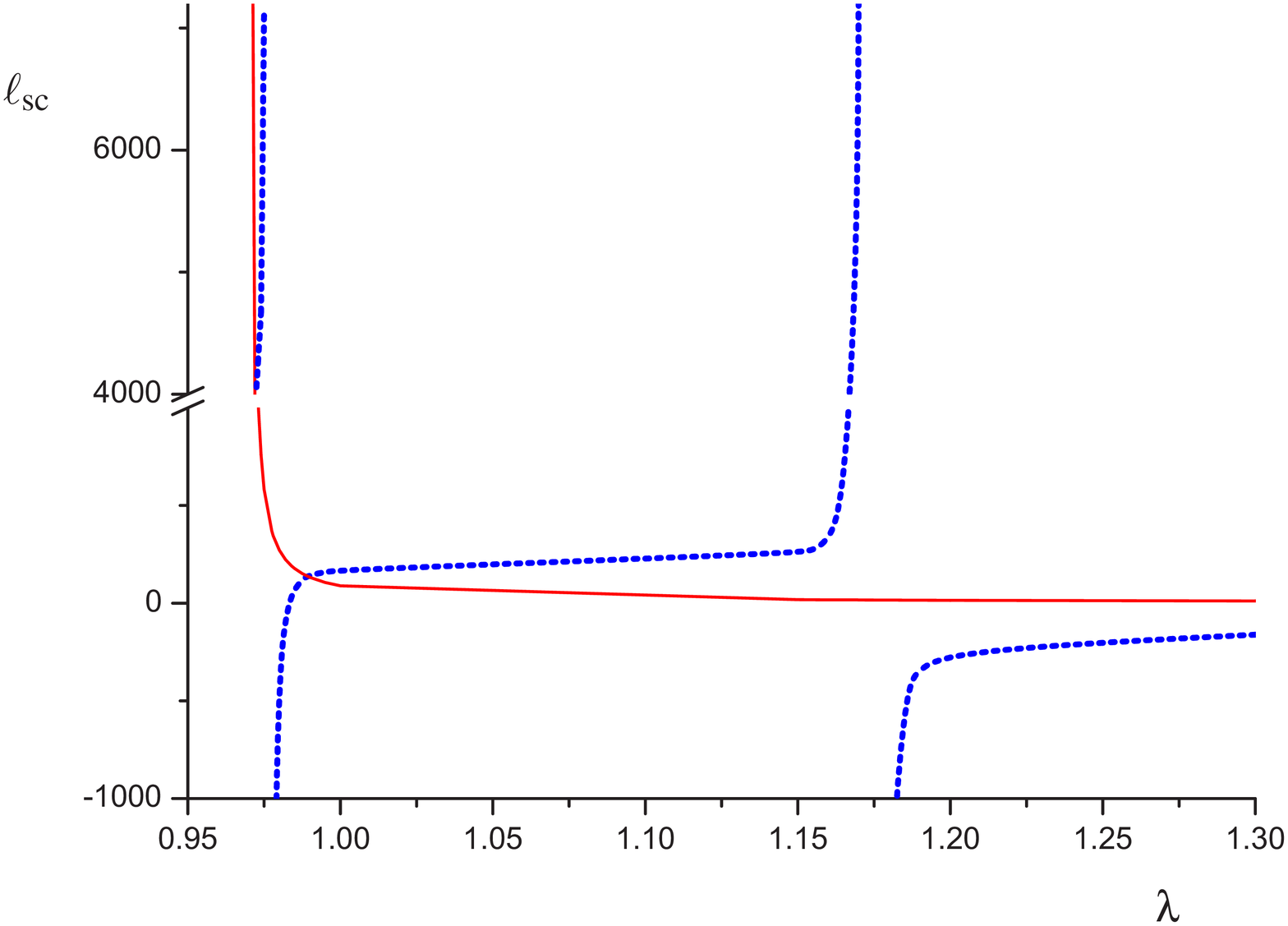}}
\vspace*{-0.7truecm}

\caption{Dependence of the atom-atom scattering length
$\ell_{\mathrm{sc}}^{(1+1)}$ (\AA) (solid curve) and atom-diatom
scattering length $\ell_{\mathrm{sc}}^{(1+2)}$ (\AA) (dashed curve)
on the factor $\lambda$.} \label{ell-lambda}
\end{minipage}
\vspace*{-0.5truecm}
\end{figure}

The third row shows values of the $^4$He--$^4$He$_2$ scattering
length. Notice that in Ref.~\cite{PRA04} our previous calculations
of  \cite{MSSK} have been improved essentially by enlarging the grid
parameters and the cut-off hyperradius. This table also contains the
results by Blume and Greene \cite{BlumeGreene}, by Ro\-ud\-nev
\cite{Roudnev1}, and the most recent ones by Lazauskas and Carbonell
\cite{Lazaus}. The treatment of \cite{BlumeGreene} is based on a
combination of the Monte Carlo method and the hyperspherical
adiabatic approach. Roudnev employs three-dimensional Faddeev
differential equations. Lazauskas and Carbonell also use
differential  Faddeev equations, but with the hard-core boundary
conditions of Ref. \cite{Vestnik}. Our results agree rather well
with these alternative calculations. For completeness we mention the
model calculations of \cite{BraatenHammer}. Being characterized by
remarkable simplicity, they rely essentially on the binding energy
obtained in  {\it ab initio} calculations.

The last two rows of  Table \ref{tableTrimers} contain results for
the binding energy of the asymmetric $^3$He$^4$He$_2$ trimer and for
the $^3$He--$^4$He$_2$ scattering length. The phase shifts
$\delta(E)$ of $^{3}$He--$^4$He$_2$ and $^{4}$He--$^4$He$_2$
scattering are depicted  in Fig. \ref{Fig-phases} where in both cases
the normalization $\delta(0) = 0$ is used.

It is already widely accepted that the excited $^4$He trimer state
is of Efimov nature. For the first time this was clearly shown by
Cornelius and Gl\"ockle \cite{Cornelius}. A more detailed
investigation has been made in \cite{YaF99} where the mechanism of
emerging new Efimov states from resonances was studied. For a latest
discussion of this subject see Ref. \cite{Lazaus}. Table
\ref{tableScLength1}, which is borrowed partly from \cite{YaF99},
demonstrates how a new Efimov level arises from a virtual state when
the interatomic interaction in the $^4$He$_3$ is being weakened. To
this end we multiply the original HFD-B potential of Ref.
\cite{Aziz87} by a factor $\lambda<1$. Decreasing this coupling
constant, there emerges, for $\lambda\approx 0.986$, a virtual state
with energy $E_\mathrm{virt}$ lying on the unphysical sheet. This
energy, relative to the diatom energy $\varepsilon_d$, is given in
the lower half of column 4. When decreasing $\lambda$ further, this
state turns at the value $\lambda\approx 0.976$ into the second
excited state. Its energy $E^{**}$ relative to $\epsilon_d$ is shown
at the bottom of the fifth column. When the second excited state
emerges, the atom--diatom scattering length $\ell^{(1+2)}_{\rm sc}$
changes its sign going through a pole, while the atom-atom
scattering length $\ell^{(1+1)}_{\rm sc}$ increases monotonically.

We have also studied the opposite case where $\lambda>1$ is
increasing. In this case the scattering length  $\ell^{(1+2)}_{\rm
sc}$ is growing until $\lambda \approx$ 1.175. There
$\ell^{(1+2)}_{\rm sc}$ becomes negative passing through a
singularity and the excited state energy $E^*$  turns into a virtual
level.  The energy $E_\mathrm{virt}$ of this state is shown
for $\lambda\geq 1.18$ in the upper quarter of column 4.

In Fig. \ref{ell-lambda} we graphically display the behavior of the
atom-atom and atom-diatom scattering lengths  $\ell^{(1+1)}_{\rm
sc}$ and $\ell^{(1+2)}_{\rm sc}$ shown in Table
\ref{tableScLength1}.


\begin{thebibliography}{9}

\bibitem{DimerExp}
F. Luo {\it et al.},
%G. C. McBane, G. Kim,  C. F. Giese,  and  W. R. Gentry,
J. Chem. Phys. \textbf{98} (1993) 3564.

\bibitem{Science}
W. Sch\"ollkopf and J. P. Toennies, Science \textbf{266} (1994)
1345.

\bibitem{exp}
R. Grisenti {\it et al.},
%W. Sch\"ollkopf, J. P. Toennies, G. C. Hegerfeld, T. K\"ohler, and M.Stoll,
Phys. Rev. Lett. \textbf{85} (2000) 2284.

\bibitem{Toennies2002} L. W. Bruch, W. Sch\"ollkopf, and J. P. Toennies,
                       J. Chem. Phys. \textbf{117} (2002), 1544.
\bibitem{Cornelius}
Th. Cornelius and W. Gl\"ockle, J. Chem. Phys. \textbf{85} (1986) 3906.

\bibitem{Nielsen}
E. Nielsen, D. V. Fedorov, and A. S. Jensen, J. Phys. B \textbf{31}
(1998) 4085.

\bibitem{BlumeGreene}
D. Blume and C. H. Greene, J. Chem. Phys. \textbf{112} (2000) 8053.

\bibitem{RoudnevYak}
V. Roudnev and S. L. Yakovlev, Chem. Phys. Lett. \textbf{328} (2000) 97.

\bibitem{Bressanini}
D. Bressanini, M. Zavaglia, M. Mella, and G. Morosi, J. Chem. Phys.
\textbf{112} (2000) 717.

\bibitem{PRA-CPL}
E. A. Kolganova, A. K. Motovilov, and S. A. Sofianos, Phys. Rev. A
\textbf{56} (1997) R1686; A.\,K.\,Motovilov, S. A. Sofianos, and E. A.
Kolganova, Chem. Phys. Lett. \textbf{275} (1997) 168.

\bibitem{MSSK}
A. K. Motovilov, W. Sandhas, S. A. Sofianos, and E. A. Kolganova,
Eur. Phys. J. D \textbf{13} (2001) 33.

 \bibitem{Barletta}
 P. Barletta and A. Kievsky, Phys. Rev. A \textbf{64} (2001) 042514.

 \bibitem{Roudnev1}
 V. Roudnev, Chem. Phys. Lett. \textbf{367} (2003) 95.

\bibitem{PRA04}
E. A. Kolganova, A. K. Motovilov, and W. Sandhas,
Phys. Rev. A \textbf{70} (2004) 052711.

\bibitem{Lazaus}
R.~Lazauskas and J.~Carbonell,
Phys. Rev. A \textbf{73} (2006) 062717.

\bibitem{FBS2004}
W. Sandhas, E. A. Kolganova, Y. K. Ho, and A. K. Motovilov,
Few-Body Syst. \textbf{34} (2004) 137.

\bibitem{Roudnev}
V. Roudnev, Private communication (2004).

\bibitem{Kraemer}T. Kraemer {\it et al.}, Nature \textbf{440} (2006) 315;
B. D. Esry and C. H. Greene, \textit{Ibid.} \textbf{440} (2006) 289.

\bibitem{BraatenHammer}
T. Frederico, L. Tomio, A. Delfino, and A. E. A. Amorim, Phys.
Rev. A \textbf{60} (1999) R9; E. Braaten and H.-W. Hammer, Phys.
Rev. A \textbf{67} (2003) 042706; F. M. Pen'kov and W. Sandhas,
Phys. Rev. A \textbf{72} (2006) 060702(R).

\bibitem{Aziz91}
R. A. Aziz and M. J. Slaman, J. Chem. Phys. \textbf{94} (1991) 8047.

\bibitem{Vestnik} A. K. Motovilov, Vestnik Leningr. univ. (fiz. khim.) \textbf{22} (1983)
76; S. P. Merkuriev, A. K. Motovilov, and S. L. Yakovlev, Theor.
Math. Phys. \textbf{94} (1993) 435.

\bibitem{Aziz87}
     R. A. Aziz, F. R. W. McCourt, and C. C. K.  Wong,
      Mol. Phys. \textbf{61} (1987) 1487.

\bibitem{YaF99} E. A. Kolganova and A. K. Motoviov, Phys. At. Nucl. \textbf{62}
(1999) 1179.

\end{thebibliography}
\end{document}